\documentclass[draft]{aipproc}

\layoutstyle{6x9}

\begin{document}

\title{Photoproduction at Hadron Colliders}


\author{Spencer R. Klein, for the STAR Collaboration}{
  address={Lawrence Berkeley National Laboratory, Berkeley, CA, 94720, USA}
}

\begin{abstract}

Photoproduction can be studied at hadron colliders by using the
virtual photons associated with the hadron beams.  The LHC will reach
photonucleon energies 10 times higher than that available elsewhere.
These reactions are already being studied at RHIC. After introducing
photoproduction at hadron colliders, I will discuss recent results
from STAR on $\rho^0$, $\pi^+\pi^-\pi^+\pi^-$ and $e^+e^-$ production.
\end{abstract}

\maketitle

\section{Introduction}

The upcoming Large Hadron Collider (LHC) will reach proton-proton
energies an order of magnitude higher than any existing accelerator.
Because relativistic protons and heavier nuclei accompanied by fields
of virtual photons, the LHC can be used to study photonuclear and
two-photon interactions at energies far beyond those accessible at
HERA or other accelerators.  

Photoproduction is of interest in both pp and heavy-ion collisions.
Proton-proton collisions produce photons with the highest energies,
and, because of the very high $pp$ luminosities, good rates.  However,
for many channels, the signal to noise ratio may be lower than in ion
collisions.  Heavy ions are accompanied by very high photon fluxes,
and, because of the very strong fields, a single ion-ion collision can
induce multiple electromagnetic interactions.  The correlations
between the multiple photons lead to the ability to ``tune'' the
photon beam, by selecting different photon energy spectra and
polarizations.  Since these reactions take place at large impact
parameters, where no hadronic interactions occur, they are often known
as ``ultra-peripheral collisions'' (UPCs).

UPCs can be used to study a variety of topics
\cite{ourreview}\cite{baurreview}\cite{kraussreview}.  Low$-x$ gluon
distributions can be probed via heavy quark (including quarkonium) and
jet production.  UPCs can be used for many other studies of
QCD.  At the LHC, photonuclear interactions can be used to search for
new physics. The strong fields allow for many tests of quantum
electrodynamics in the very strong field regime, where perturbation
theory may be expected to fail.  Many of these topics are already
being studied at RHIC.

\section{Photoproduction at Hadron Colliders}

For most reactions, the photon flux from protons or nuclei is well
described by the Weizs\"acker-Williams method of virtual photons.  The
photon flux per unit area for an energy $\omega$ at a distance $b$
from a relativistic nucleus with charge $Z$ is \cite{jackson}
\begin{equation}
N(\omega,b)= \frac{Z^2\alpha\omega^2}{\pi^2\gamma^2\hbar^2}
K_1^2(x)
\label{eq:fflux}
\end{equation}
where $x=\omega b/\gamma$, $\gamma$ is the Lorentz boost of the
nucleus, $\alpha\approx 1/137$ is the fine structure constant, and
$K_1$ is a modified Bessel function.  The total photon flux from an
ion with radius $R_A$ is
\begin{equation}
n(\omega) = \int d^2b N(\omega,b).
\label{eq:ntot}
\end{equation}
The constrant $b>R_A$ is usually imposed to eliminate the photon flux
inside the nucleus (where Eq. \ref{eq:fflux} fails, and, in any case,
most of the flux is not usable).  For photonuclear or two-photon
interactions to be visible, the two nuclei must not interact
hadronically, requiring $b>2R_A$.  This flux is calculated
numerically, but can be approximated within about 15\% by
requiring $b>2R_A$ in Eq. \ref{eq:ntot} \cite{ourreview}.

The cross section for photonuclear interactions can be written
\cite{BKN}
\begin{equation}
\sigma(A+A\rightarrow A+A+X) = \int d^2b P(b)
\end{equation}
where $P(b)$ is the probability for a photonuclear interaction,
$P(b)=\int d\omega N(\omega,b)\sigma_{\gamma A}(\omega)$ where
$\sigma_{\gamma A}(\omega)$ is the cross section for the photonuclear
interaction in question.  This formulation is easy to generalize to
include multiple interactions between a single ion pair:
\begin{equation}
\sigma (A + A\rightarrow X_1 + X_2+...) = \int d^2 P_1(b)P_2(b).
\end{equation}
In general, $P(b)\approx 1/b^2$, so the integrand for a $n-$photon
reaction goes as $1/b^{2n}$ and the more photons involved in a
reaction, the smaller the average impact parameters \cite{factorize}.
For example, with gold at RHIC, the median impact parameter drops from
46~fm for unselected $\rho^0$ production to 18 fm for $\rho^0$
accompanied by mutual Coulomb excitation \cite{BKN}.  The smaller
impact parameters harden the photon spectrum, from $1/\omega$ to
independent of $\omega$.  For some reactions, $P(2R_A)>1$;
in this case $P(b)$ is the mean number of reaction at that $b$.

Factorization can be used to simplify triggering on UPCs.  One
reaction can serve as a 'trigger' for another.  STAR has studied the
reactions $Au + Au\rightarrow Au^* + Au^* + \rho^0$ and $Au +
Au\rightarrow Au^* + Au^* + e^+e^-$, using signals from the neutrons
emitted in the $Au^*$ decays to trigger the detector, providing
$\rho^0$ and $e^+e^-$ samples without trigger bias.

The photon polarizations are also correlated.  The electric field of
the photon-emitting nucleus parallels the impact parameter vector, so
photons are linearly polarized along the impact parameter vector.  For
multiple interactions between a single ion-pair, the parallel
polarizations can lead to observable angular correlations between
decay products \cite{factorize}.

\section{Results from STAR at RHIC}

The STAR collaboration has produced final results on $\rho^0$
production \cite{STARrho} and on two-photon production of $e^+e^-$
pairs \cite{STARee}.  Events were selected with two types of triggers:
minimum bias triggers that select events with mutual Coulomb
dissociation, taking advantage of factorization, and topological
triggers, that select low multiplicity events with appropriate
topologies in the central detector \cite{STARrho}.

The $\rho$ data is well described by the soft Pomeron model, and the
previously discussed factorization holds.  In the soft Pomeron model,
the incident photon fluctuates to a quark-antiquark pair, which then
elastically scatters (via Pomeron exchange) from the target nucleus
\cite{KN99}.  Because the scattering is coherent, the momentum
transfer is limited to order $\hbar/R_A$.  This low $p_T$ is a
distinctive experimental signature; for gold, most of the
signal occurs for $p_T<150$ MeV/c.

\begin{figure}[tb]
\resizebox{0.8\columnwidth}{!}
{\includegraphics[clip]{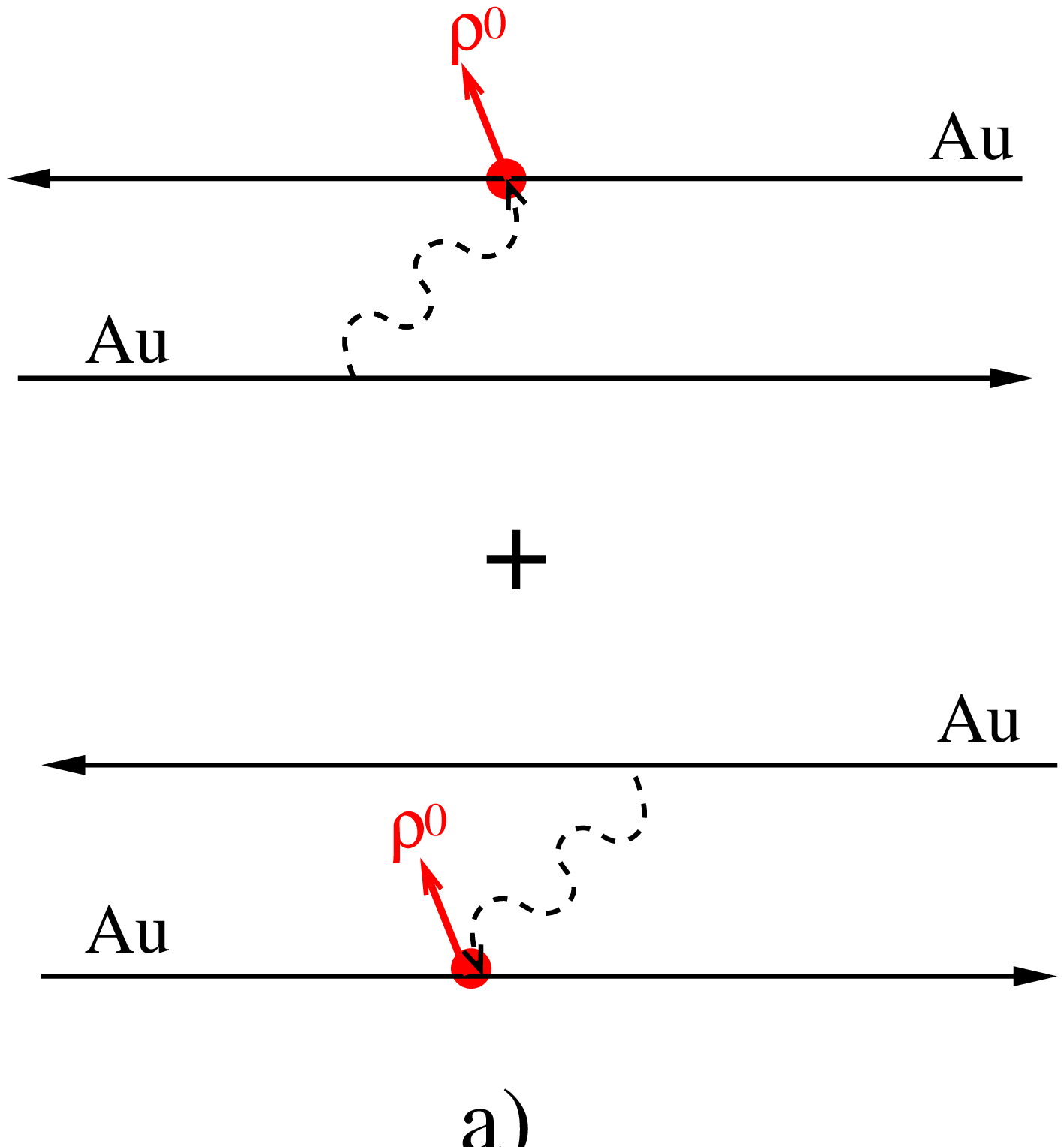}
\hskip 0.1 in 
\includegraphics[clip,scale=1.2]{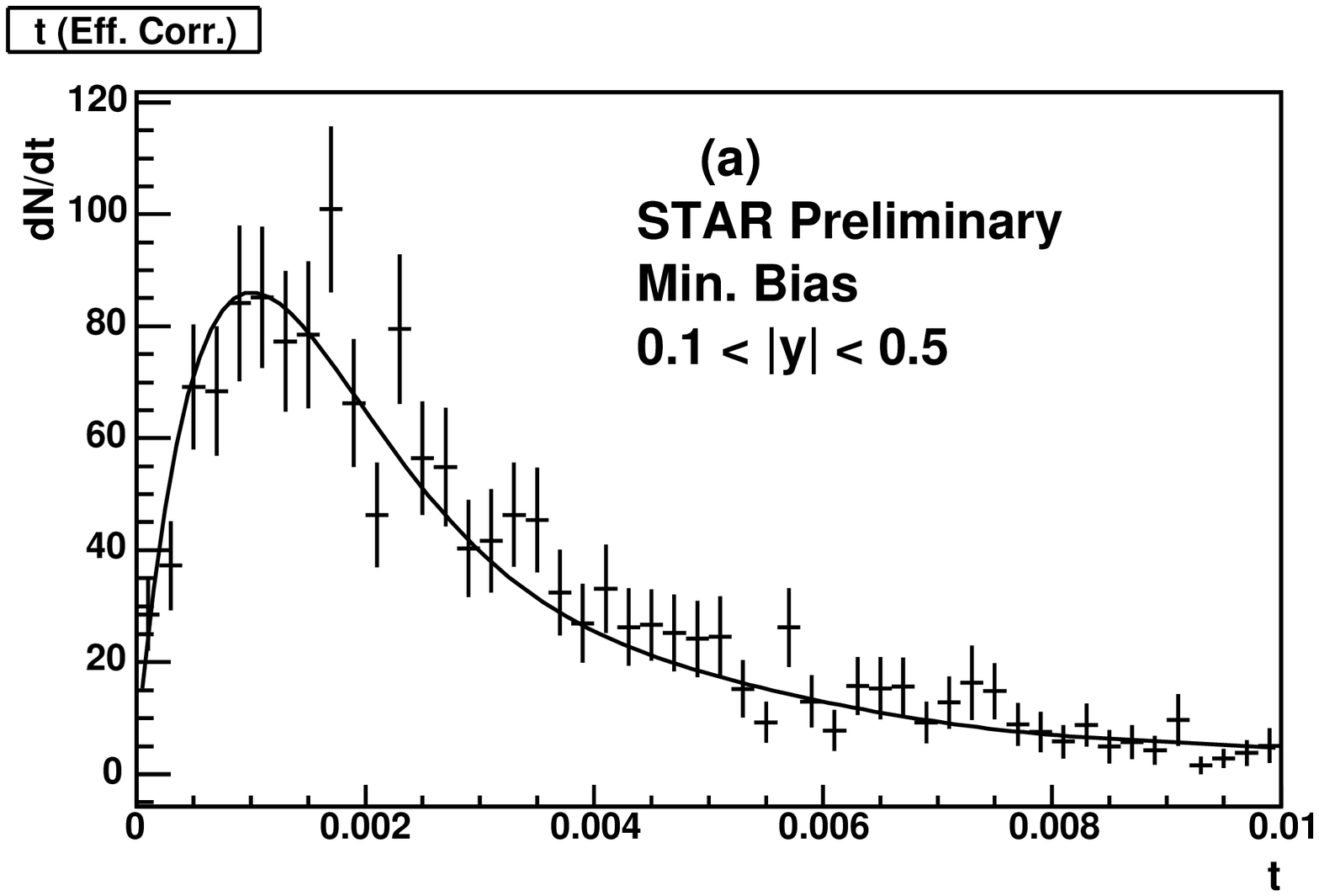}}
\caption{(a) Diagrams for the two interferening production mechanisms,
at two spatially separated locations. (b) $t_\perp = p_T^2$ spectrum
for $\rho^0$ accompanied by mutual Coulomb excitation.  The drop at
low $t_\perp$ is due to interference.  The solid histogram is a fit to
the data, using a model that includes the interference.}
\end{figure}

Figure 1 shows the $t_\perp = p_T^2$ spectrum of $\rho^0$ with
rapidity $0.1 < |\eta| < 0.6$, selected with stringent cuts to
minimize the background \cite{STARinterfere}.  At moderate and high
$t$, the spectrum is well fit by an exponential, $dN/dt = a
\exp{(-bt)}$.  However, for $t<0.0015$GeV$^2$, the data drops off.
This drop can be explained by interference between two
indisinguishable possibilities: nucleus 1 emits a photon which
interacts with nucleus 2, or vice-versa \cite{theoryinterfere}.  In
$pp$ or $AA$ collisions, these two possibilities are related by a
parity transformation.  Since the $\rho^0$ is negative parity, the
interference is destructive.  At mid-rapidity,
\begin{equation}
\sigma = \sigma_0 \big[1-\cos{(p_Tb)}\big]
\label{eq:interf}
\end{equation}
Of course, $b$ is unknown, and the overall interference depends on the
integral over all $b$.  Away from $y=0$, the interference is reduced
because the photon energies, fluxes, amplitudes {\it etc.} for the two
directions are different.  The solid curve in Fig. 1 shows a fit to a
functional form based on these factors; for this sample, the
interference is $101\pm8 {\rm (stat.)} \pm15 {\rm (syst.)} \%$ of that
expected \cite{STARinterfere}.  Because the two sources are spatially
separated, the final state $\pi^+\pi^-$ wave function does not
factorize into single-particle wave functions, and the system exhibits
the Einstein-Podolsky-Rosen paradox \cite{EPR}.  For $\overline p p$
collisions, the transformation between the two possibilies is a
charge-parity transformation; vector mesons are $CP$ positive, so the
interference in Eq.~\ref{eq:interf} is positive \cite{ppinterfere};
this may be studied at the Fermilab Tevatron.

STAR has also studied $\rho^0$ production in $dAu$ collisions.  The
photon is usually emitted by the gold nucleus, and the deuteron is the
target.  Both coherent (deuteron stays intact) and incohent (deuteron
dissociates) interactions have been observed.  The $t_\perp$ spectrum
for the incoherent interactions is similar to that observed in $eA$
collisions at HERA \cite{timser}.

The STAR $e^+e^-$ data is well described by lowest order quantum
electrodynamics and factorization \cite{STARee}.  The $p_T$
spectrum of the $e^+e^-$ pairs is not well described by the virtual
photon paradigm - the photon virtuality is required
to fit the data.

STAR has also studied 4-prong final states, like
$\pi^+\pi^-\pi^+\pi^-$.  Fig. 2 compares the $p_T$ spectrum of 4-prong
events with net charge 0 with those of net charge 2.  This data was
taken in 2002 with the minimum-bias trigger.  A neutral excess is
present for $p_T < 150$ MeV/c, with the mass spectrum of the excess
centered around 1.5 GeV/c$^2$.

\begin{figure}[tb]
\resizebox{0.7\columnwidth}{!}
{\includegraphics[clip,height=1.5in]{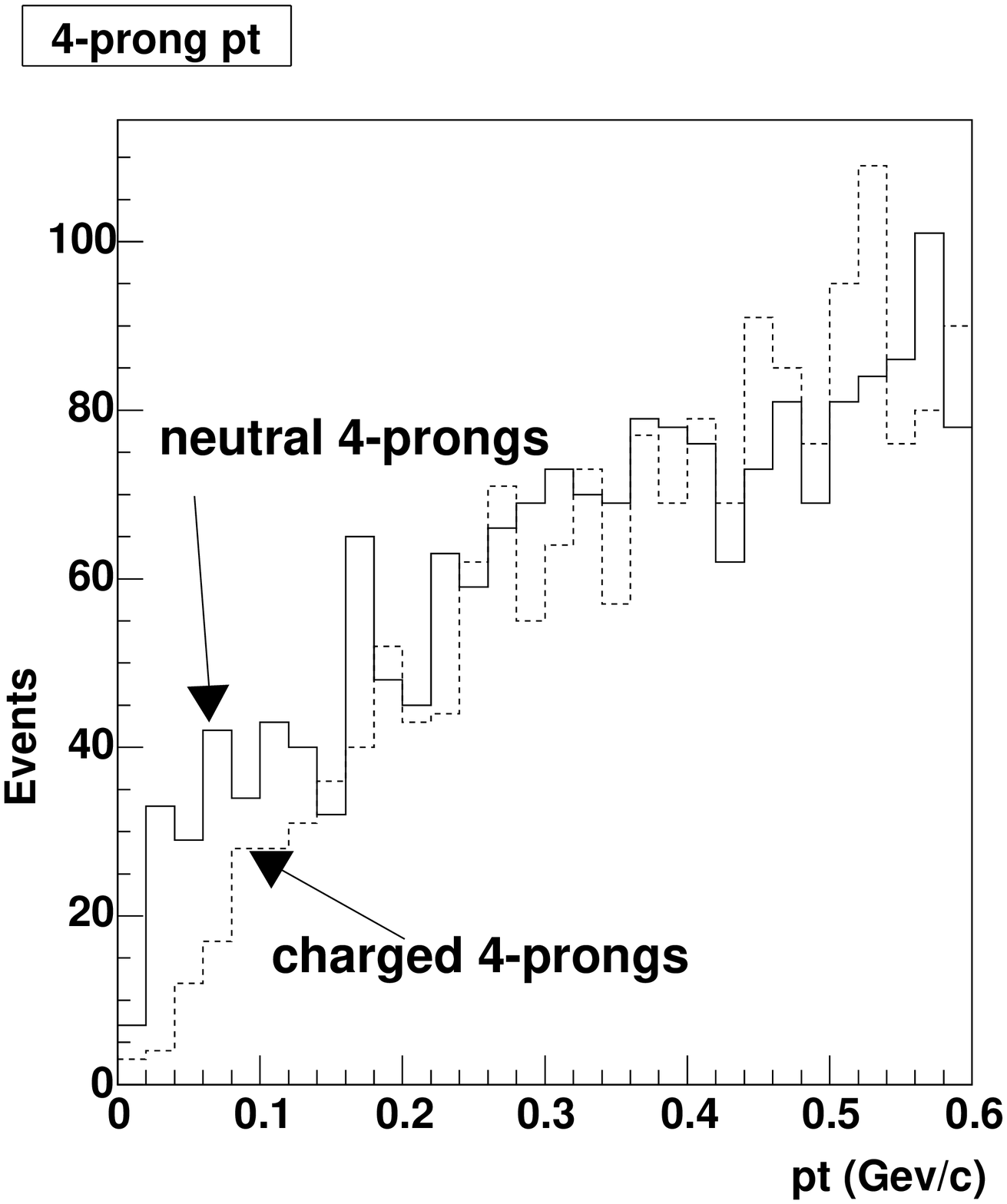}
\includegraphics[clip,height=1.5in]{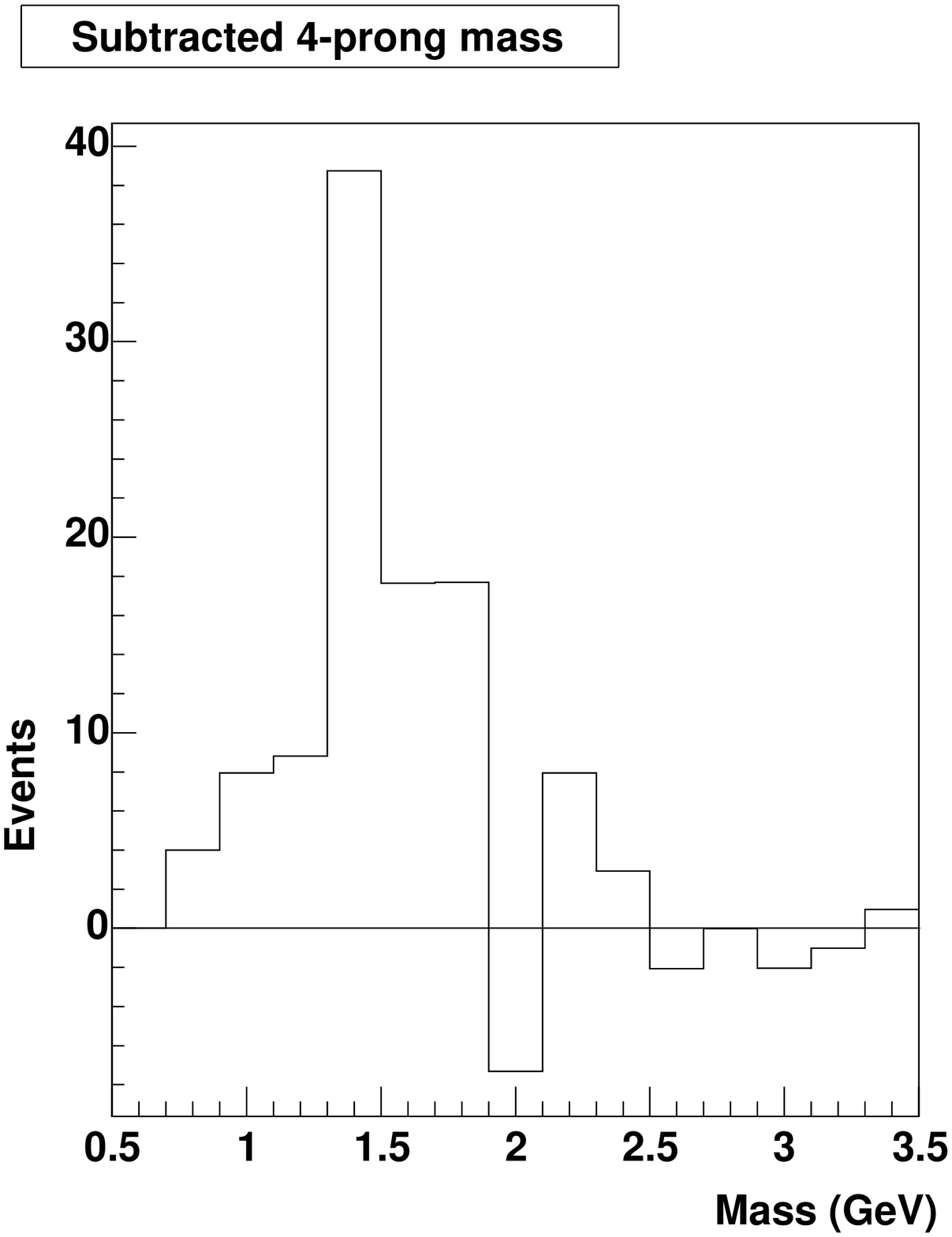}}
\caption{(a)$p_T$ spectrum ($dN/dp_T$) for 4-prong events for neutral
and net-charged combinations. (b) The mass spectrum for the neutral
excess is peaked around 1.5 GeV.}
\end{figure}

This work was supported by the U.S. D.O.E. under Contract No.
DE-AC-03076SF00098.

\end{document}